\documentclass[11pt,letterpaper]{article}

\usepackage[T1]{fontenc}
\usepackage[utf8]{inputenc}
\usepackage[margin=1in]{geometry}
\usepackage{graphicx}
\usepackage{amsmath,amssymb}
\usepackage{newtxtext,newtxmath}
\usepackage{array,booktabs}
\usepackage{placeins}
\usepackage{microtype}
\usepackage{xurl}
\usepackage[numbers,sort&compress]{natbib}
\usepackage{fancyhdr}
\usepackage{hyperref}
\hypersetup{hypertexnames=false,colorlinks=true,linkcolor=black,citecolor=black,urlcolor=blue,bookmarksdepth=2,
pdftitle={Patch correctness in automated program repair: A five-dimensional evidence framework and cross-benchmark audit from classical repair to coding agents},
pdfauthor={Shuai Wu, Yanna Feng, Yufang Li, Xue Li, Zhijun Wang, Bolun Liu, Weilin Cai}}
\fancypagestyle{titlepage}{%
  \fancyhf{}%
  \fancyfoot[L]{\footnotesize\begin{minipage}[b]{\textwidth}\raggedright
    *Corresponding author: Yanna Feng, \href{mailto:yanna.feng@outlook.com}{yanna.feng@outlook.com}.\\
    ORCID: Shuai Wu \href{https://orcid.org/0009-0007-7657-6208}{0009-0007-7657-6208};
    Yanna Feng \href{https://orcid.org/0009-0000-6443-6178}{0009-0000-6443-6178};
    Xue Li \href{https://orcid.org/0009-0002-7480-723X}{0009-0002-7480-723X};
    Zhijun Wang \href{https://orcid.org/0009-0000-0742-6985}{0009-0000-0742-6985}.
    \end{minipage}}%
}

\newcommand{\authorcell}[2]{%
  \begin{minipage}[t]{0.22\textwidth}\centering
  \textbf{#1}\par\smallskip
  {\small #2}\par
  \end{minipage}%
}
\newcommand{\authorcellb}[2]{%
  \begin{minipage}[t]{0.30\textwidth}\centering
  \textbf{#1}\par\smallskip
  {\small #2}\par
  \end{minipage}%
}

\begin{document}
\thispagestyle{titlepage}
\begin{center}
{\small\scshape Technical Report\par}\vspace{0.4em}
\rule{\textwidth}{2pt}\par\vspace{0.7em}
{\LARGE\scshape Patch correctness in automated program repair:\\A five-dimensional evidence framework and cross-benchmark audit from classical repair to coding agents\par}
\vspace{0.7em}\rule{\textwidth}{2pt}\par
\end{center}

\vspace{1.1em}
\noindent\makebox[\textwidth][c]{%
\authorcell{Shuai Wu}{University of Colorado Boulder}\hspace{0.03\textwidth}%
\authorcell{Yanna Feng\textsuperscript{*}}{Beijing Union University}\hspace{0.03\textwidth}%
\authorcell{Yufang Li}{Beijing Union University}\hspace{0.03\textwidth}%
\authorcell{Xue Li}{Beijing Union University}}\par
\vspace{1.3em}
\noindent\makebox[\textwidth][c]{%
\authorcellb{Zhijun Wang}{University of North Dakota}\hspace{0.03\textwidth}%
\authorcellb{Bolun Liu}{City University of Macau}\hspace{0.03\textwidth}%
\authorcellb{Weilin Cai}{The University of Sydney}}\par

\vspace{1.2em}
\begin{center}November 2026\end{center}

\begin{abstract}
Tests guide automated program repair and determine which candidate patches are reported as successful. Repository-level coding agents extend this feedback loop by generating tests and revising code after observing execution results. This technical report introduces a five-dimensional evidence profile and reporting protocol to audit how test observations support patch-correctness claims across classical repair systems and coding agents. We assemble an audit dataset of 39 benchmark and evaluation artifacts: 32 primary empirical studies, one institutional technical audit, and six prior taxonomic baselines. We code the 33 primary contributions along five dimensions: test provenance and exposure, behavioral scope, oracle basis and dependence, execution integrity, and label adjudication. The central distinction separates checks visible during repair search from evidence reserved for final assessment. Four worked comparative audits show how changed candidate populations, oracle rules, or execution budgets alter reported detection rates. Additional tests can expose missed faults while rejecting valid alternatives or reducing correct-repair yield under a fixed budget. Repository-level audits separate issue reproduction from candidate-patch discrimination and distinguish automated discrepancy detection from requirements-based adjudication. An evidence matrix links each coded decision to supporting passages and records shared benchmark ancestry. We specify a reporting protocol for repair benchmarks and maintenance pipelines that retains uncertain judgments and records validation cost alongside fault detection and valid-patch retention.
\end{abstract}

\bigskip

\section{Introduction}\label{sec:introduction}

A software patch can satisfy every test used during its construction and still violate an intended maintenance requirement. This problem has shaped automated program repair (APR) for more than a decade. Early empirical studies examined whether repairs generalized beyond their guiding test suites and whether accepted patches preserved existing functionality. Their findings made test-suite overfitting a central concern when interpreting repair results \citep{smith2015,qi2015}. Repository-level systems face the same evidential problem in a larger execution context. A single patch may modify several files, depend on a specific container environment, and emerge from an iterative sequence of test executions and code revisions.

SWE-bench introduced a practical issue-resolution setting in which a system receives an issue description and a repository snapshot, then produces a patch evaluated against task-associated tests \citep{jimenez2024swebench}. SWE-agent exposes repository navigation, file editing, and test execution through an agent-computer interface \citep{yang2024sweagent}. Agentless separates localization, repair, and validation into a staged pipeline and uses generated reproduction tests when developer tests are withheld \citep{xia2025agentless}. Both workflows make test feedback part of the mechanism that selects a patch. An evaluation therefore needs to record which tests influenced candidate selection and which evidence was collected only after selection finished.

A benchmark score may refer to passage of selected tests, passage of a broader regression suite, agreement with a reference implementation on generated inputs, or a human judgment about software requirements. Each observation answers a different validation question, yet labels such as ``correct'' or ``resolved'' often collapse these distinctions. Prior taxonomic work on patch correctness assessment catalogues validation algorithms and their datasets \citep{fei2025}, and a separate analysis of APR benchmarks documents variation in correctness definitions, localization assumptions, stopping rules, and training-data exposure \citep{renzullo2025}. These findings motivate a structured audit of how test evidence accumulates inside classical and agentic repair workflows.

This technical report formalizes the relationship between a patch-correctness claim and the test evidence supporting it. Our unit of analysis is an evaluation decision about a candidate patch or a test suite under stated execution conditions. The same analytical record connects classical overfitting experiments with repository-level audits of generated tests, broader regression execution, reference-based discrepancy detection, and candidate-patch discrimination.

We apply a five-dimensional evidence profile to 33 primary benchmark and validation studies to analyze four recurring comparison problems across classical and agentic repair. The profile records test provenance and exposure, behavioral scope, oracle basis and dependence, execution integrity, and label adjudication. A newly generated test can extend behavioral coverage while sharing the issue interpretation that produced the broken patch. The profile also separates training contamination from test-suite overfitting: contamination concerns prior exposure to benchmark artifacts, whereas overfitting concerns generalization beyond the checks that accepted a patch. Both affect evaluation decisions and require separate evidence.

Our contribution is a methodological measurement framework and cross-benchmark audit. We do not estimate a single pooled prevalence of incorrect patches across the APR literature; reported rates remain tied to their original candidate samples and validation procedures. The five-dimensional profile structures the 33-study evidence matrix, four worked comparative audits, and a practical reporting protocol for benchmark designers and repair pipelines. Its inter-annotator reliability and operational effect on downstream review effort require prospective empirical testing.

\section{Audit methodology and dataset construction}\label{sec:design}

\subsection{Research questions and analytical boundaries}\label{sec:questions-and-boundaries}

This study audits test-based assessment of source-code patches, from classical generate-and-validate repair to systems using large language models (LLMs) and repository-level coding agents. Studies of function-level code generation are included where they directly examine test strength or temporal exposure, and their results are kept separate from repository-repair evidence. General agent planning, model architecture, and security benchmarks fall outside our scope unless they alter the interpretation of patch validation.

The audit addresses four research questions:

\begin{itemize}
\item \textbf{RQ1:} What behavioral claims are supported by reproduction tests, regression tests, differential tests, and patch-discrimination tests?
\item \textbf{RQ2:} How do test availability, iterative feedback, and training exposure affect the independence of validation?
\item \textbf{RQ3:} What do empirical audits show about false acceptance, false rejection, and uncertainty in correctness labels?
\item \textbf{RQ4:} What should an evaluation report so that its patch-correctness claims can be checked and compared?
\end{itemize}

\subsection{Corpus construction and artifact selection}\label{sec:discovery-and-selection}

We executed 46 targeted search queries on 25 October 2026 across four technical workstreams: 16 on classical repair and patch assessment, 18 on repository-level coding agents and test benchmarks, three on prior taxonomic baselines, and nine on supplementary counterevidence, of which eight completed and one encountered a transport error, followed by 24 targeted full-text and section retrievals from publisher pages, proceedings, author-hosted manuscripts, and versioned arXiv records. An additional bibliographic cross-check was performed for the benchmark study by \citet{renzullo2025}. Query strings covered patch correctness assessment, test-suite overfitting, repository-level repair, generated reproduction tests, differential testing, and benchmark contamination. Named-paper queries resolved foundational studies and publication versions, and reference lists in prior baselines supplied further candidates. The archived search log records all 46 query strings and requested result counts; those counts describe retrieval requests rather than retained studies.

Artifact selection targeted concrete validation mechanisms. We retained a source when it supplied an operational validation procedure, empirical counts on accepted patches, or an explicit audit of a benchmark assumption. Prior taxonomic baselines were retained to position the coding schema, while original empirical claims were traced to primary studies. Papers that only introduced a repair tool or restated a leaderboard score without validation analysis were excluded. After assembling an initial 31-source dataset, we conducted a supplementary coverage audit targeting contrary findings, valid alternative implementations, test-parser defects, and the downstream cost of patch filtering. The archived candidate ledger records 57 evaluated records: 32 duplicate manifestations of already-retained or newly matched studies, 15 exclusions with documented reasons, two late-October 2026 preprints held pending because publication availability by 25 October 2026 was unverified, and eight newly added sources \citep{yang2017opad,yu2019unsat,wang2019different,yu2025utboost,zemin2025criteria,song2025prism,yu2026sweabs,openai2026audit}. The resulting dataset comprises 39 sources: 32 primary empirical studies, one institutional technical audit treated as primary grey literature \citep{openai2026audit}, and six prior taxonomic baselines.

This corpus construction supports a comparative audit of validation mechanisms and evaluation assumptions. Because artifact collection targeted distinct validation designs rather than a census of subscription bibliographic databases, we do not use publication counts to infer longitudinal trends or pool effect sizes across incompatible denominators.

\subsection{Evidence coding and denominator normalization}\label{sec:extraction-and-synthesis}

For each primary contribution, we extracted the task population, source and visibility of tests, program versions compared, definition of success, numerical denominators, handling of execution errors, and procedure for assigning correctness labels. We separated directly observed failures from extrapolations based on inspected samples. Preprints are identified in the reference list, and version-specific counts are retained where metadata or numerical summaries differ across releases. When both a peer-reviewed publication and an earlier preprint were accessible, we extracted from the peer-reviewed version.

We conducted artifact collection, passage extraction, five-dimensional evidence coding, and denominator verification against the primary source texts. Coding entries and numerical denominators were cross-checked by the authors against section and table locators without a separate multi-rater blinding protocol, so inter-annotator agreement statistics are not reported.

We coded the 33 primary contributions, comprising 25 direct patch-assessment or benchmark-audit studies and eight supporting benchmark or repair-system studies, along the five dimensions in Section~\ref{sec:profile}; the six taxonomic baselines remain in the source inventory without being double-counted as primary observations. Each matrix cell records a descriptive category, a supporting passage summary, and a section or table locator. The coding rules in Appendix~\ref{app:coding} separate a method's runtime oracle from the reference labels used to evaluate that method, and distinguish an unreported control from evidence that the control was absent. Shared benchmarks and inherited label sets are recorded as dependency groups that trace dataset ancestry.

Our comparative analysis evaluates the assumptions of each validation procedure and the counterevidence it can produce. Numerical results from different candidate populations are kept separate. For example, a proportion calculated among patches already passing generated tests has a different denominator from a task-level resolution rate, and a differential-testing audit may inspect a sample enriched for suspicious patches. These denominator distinctions are preserved throughout our analysis.

\subsection{Comparison with prior evaluation taxonomies}\label{sec:relationship-to-existing-reviews}

The prior taxonomic baselines in Table~\ref{tab:reviews} provide several distinctions that we build on. Fei et al. separate methods that require oracle patches from those that operate without them, and catalogue assessment techniques, label construction, datasets, and metrics \citep{fei2025}. Renzullo et al. examine how machine learning challenges benchmark assumptions \citep{renzullo2025}. Broader LLM repair taxonomies classify model architectures, repair scenarios, and control over the repair loop \citep{zhang2026review,yang2025review}, while a life-cycle benchmark taxonomy maps code benchmarks across development activities \citep{wang2026review}. Unlike prior work that catalogues assessment algorithms or benchmark families in isolation, this technical report applies a unified decision-level evidence record across both classical APR and agentic repair loops, linking reported detection rates to candidate denominators, oracle dependencies, valid-patch retention, and downstream bug-repair yield.

\begin{table}[!htbp]
\centering\small
\caption{Comparison of prior taxonomic baselines and the analytical focus of this technical report.}
\label{tab:reviews}
\begin{tabular}{@{}>{\raggedright\arraybackslash}p{.24\linewidth}>{\raggedright\arraybackslash}p{.34\linewidth}>{\raggedright\arraybackslash}p{.36\linewidth}@{}}
\toprule
\textbf{Baseline study} & \textbf{Main organizing question} & \textbf{Connection to the present audit} \\
\midrule
\citet{fei2025} & How is patch correctness assessed? & Classifies oracle-patch dependence, assessment methods, label construction, and metrics. \\
\citet{renzullo2025} & How do emerging APR techniques challenge benchmarks? & Documents evaluation inconsistencies and training-exposure concerns. \\
\citet{zhang2026review} & How are LLMs used in APR? & Covers models, repair scenarios, and integration practices. \\
\citet{yang2025review} & Who controls the repair loop and how are models adapted? & Distinguishes procedural and agentic use of test feedback. \\
\citet{wang2026review} & Which development activities do code benchmarks measure? & Places repair evidence within broader benchmark coverage. \\
Present report & What evidence supports a particular patch-correctness claim? & Applies a five-dimensional decision record across classical and agentic studies to test what reported scores show. \\
\bottomrule
\end{tabular}
\end{table}

\section{What a passing test shows}\label{sec:claims}

\subsection{Plausibility, correctness, and the target requirement}\label{sec:plausibility-correctness-and-the-target-requirement}

A test-based evaluation observes whether a candidate patch produces the expected outcomes on a specified test suite. Correctness with respect to a software requirement is a broader claim: the patched program must satisfy intended behaviors across its operating domain. A finite test suite samples only part of that domain. Inferring requirement satisfaction from test passage depends on three factors: the behaviors covered by the suite, the validity of its expected outcomes, and the execution environment \citep{barr2015,fei2025}.

Plausibility denotes acceptance by a specified test suite. A patch may be plausible under the benchmark suite and rejected by a broader regression suite. A patch judged correct under a particular requirement can also differ from the developer's implementation. When reference code supplies expected results, it provides evidence about intended behavior, although its incidental implementation choices may be irrelevant to the issue. In a study of 177 patches judged correct, 45 differed syntactically from the developer patches, and some alternatives changed different locations \citep{wang2019different}. These findings challenge exact textual matching as a correctness rule. Whether behavioral differences are permissible requires a separate requirement-level judgment.

The test-oracle problem concerns how an expected outcome is determined \citep{barr2015}. In program repair, the oracle may be an existing assertion, a newly written requirement test, a reference implementation, or a human reviewer's judgment. An evaluation should state which source supplies the verdict. Increasing the number of test executions expands behavioral sampling, yet it does not by itself verify every expected outcome.

\subsection{Reproduction and regression preservation}\label{sec:reproduction-and-regression-preservation}

A reproduction test typically fails on the buggy version and passes after applying a reference fix. This fail-to-pass transition shows that the test distinguishes those two program states. SWT-Bench and TDD-Bench Verified use this transition to evaluate issue-related test generation, supplemented by measures such as coverage of modified code \citep{mundler2024swtbench,ahmed2024tddbench}. The criterion filters out tests that pass both versions without exposing the target failure.

The transition leaves several questions open. The test may encode only one example of the intended requirement. It may also depend on a reference-specific interface or message, and a different implementation may satisfy the underlying issue while failing the test. Conversely, a patch can reproduce the reference's behavior on this example while breaking functionality elsewhere. Regression tests address the latter concern by checking behaviors expected to remain stable. Reproduction and regression preservation should therefore be reported as separate observations.

Coverage can show whether a test reached changed lines, branches, or functions. The associated assertions determine whether an incorrect output would be detected: a test can execute a changed statement without checking the affected behavior. TDD-Bench reports the fail-to-pass condition and changed-code coverage separately \citep{ahmed2024tddbench}.

\subsection{Differential testing and alternative implementations}\label{sec:differential-testing-and-alternative-implementations}

Differential testing compares candidate and reference executions on common inputs. A discrepancy can reveal a missed boundary case or an unintended change. The evidence is strongest when the input satisfies the task's preconditions and the expected behavior follows from the requirement. Where these conditions are uncertain, discrepancy supplies a focused question for further inspection.

Reference-based checking can also reject a valid alternative implementation. Two patches may return different permissible outputs, expose equivalent behavior through different internal structures, or handle unspecified inputs differently. An evaluation therefore needs to separate behavioral divergence from adjudicated incorrectness. The PatchDiff study makes this separation explicit by combining automated discrepancy detection with manual inspection and an uncertain label category \citep{wang2026patchdiff}. Earlier reference-based test-generation work similarly documented false alarms caused by generated inputs and reference-specific behavior \citep{ye2021rgt}.

\subsection{A five-dimensional evidence profile}\label{sec:profile}

\begin{figure}[!htbp]
\centering
\includegraphics[width=\linewidth]{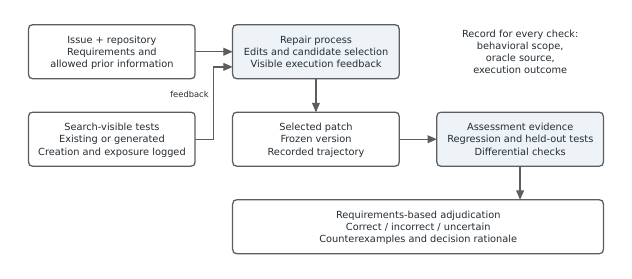}
\caption{Information flow in test-based repair assessment. Search-visible tests and feedback affect candidate selection. Independent checks add evidence with their own behavioral and oracle assumptions. Adjudication relates observations to requirements and retains unresolved cases.}
\label{fig:profile}
\end{figure}

Figure~\ref{fig:profile} separates evidence used during repair from evidence collected for final assessment. The proposed profile records five dimensions for every reported judgment. These dimensions are descriptive; they are not a numerical quality score. Strong evidence on one dimension does not automatically compensate for a weakness on another.

\begin{table}[!htbp]
\centering\small
\caption{Evidence profile for a patch-validation decision.}
\label{tab:profile}
\begin{tabular}{@{}>{\raggedright\arraybackslash}p{.24\linewidth}>{\raggedright\arraybackslash}p{.34\linewidth}>{\raggedright\arraybackslash}p{.36\linewidth}@{}}
\toprule
\textbf{Dimension} & \textbf{Information to record} & \textbf{Interpretation it supports} \\
\midrule
Test provenance and exposure & Test origin; creation time; prompts and source artifacts; visibility of tests, results, and reference code during repair. & Identifies which observations influenced candidate selection and which were held out. \\
Behavioral scope & Reproduction cases; regression targets; input domains; properties; changed components; candidate-patch diversity. & Defines the behaviors sampled and the claim's applicable scope. \\
Oracle basis and dependence & Source of expected outputs; shared models and prompts; reference dependence; training-data exposure. & Identifies dependencies between patch generation and its judge. \\
Execution integrity & Repository and dependency versions; harness; selected tests; skips, errors, timeouts, and repeated executions. & Determines what was actually tested and which outcomes can be reproduced. \\
Label adjudication & Decision rules; supporting counterexamples; reviewer process; disagreements and uncertain cases. & Distinguishes observed test outcomes from judgments about requirements. \\
\bottomrule
\end{tabular}
\end{table}

The profile keeps several kinds of independence separate. Withholding both test content and outcomes until candidate selection is complete provides procedural separation for final assessment. Generating the test with the same model may preserve an interpretive dependency. A hidden developer test may also concern an artifact present in pretraining. Each situation has a different evidential meaning, so a single ``independent validation'' flag would discard useful information.

\section{Lessons from classical program repair}\label{sec:classical}

\subsection{Acceptance depends on both tests and their harness}\label{sec:acceptance-depends-on-both-tests-and-their-harness}

Early generate-and-validate systems made test execution the main acceptance mechanism. GenProg's empirical evaluation illustrates this setting: positive and negative tests guide a search over program variants \citep{legoues2012}. Subsequent audits show why the meaning of acceptance needs to include the test harness. Qi et al. re-executed archived patches and found failures in the intended validation outcomes as well as patches that passed the tests while violating functionality \citep{qi2015}. Some harnesses used weak output proxies, and the PHP setup allowed the patched interpreter to participate in judging its own tests.

Misinterpreting a test result undermines even the basic claim that a patch passes the specified suite. Accurate execution of an incomplete suite shows test-suite plausibility while leaving untested behaviors open. Contemporary repair evaluations must verify both the execution harness and the test suite. An exit status is informative only when it faithfully reflects the intended assertions. The harness should retain collected test identifiers, per-test outcomes, and diagnostic logs so that skipped tests or parser errors can be separated from completed passing runs.

Smith et al. investigated generalization by separating the tests used for repair from independently developed evaluation tests \citep{smith2015}. Their findings support a relationship between stronger construction suites and improved held-out performance within the studied small-program setting. They also found overfitting in novice human patches. Human authorship therefore identifies provenance, while correctness requires behavioral evidence. Patch minimization alone did not eliminate the problem.

\subsection{Candidate generation and candidate selection interact}\label{sec:candidate-generation-and-candidate-selection-interact}

Long and Rinard analyzed search spaces for SPR and Prophet and found that plausible incorrect candidates outnumbered correct patches by orders of magnitude across the studied defects \citep{long2016}. Expanding a search space altered both validation cost and the order in which candidates were encountered. Under stopping policies that accept an early plausible patch, a larger search space reduced the discovery of correct repairs in their experiments.

When an LLM samples many candidate patches, a selector has more opportunities both to find a valid fix and to encounter a patch that exploits an incomplete judge. The quantitative ratios observed with classical search tools remain specific to those systems, so generation and validation budgets should be reported together. Evaluating the best candidate selected from a large pool answers a different question from evaluating the first plausible patch returned under a fixed operational budget.

Benchmark granularity also shapes validation evidence. Defects4J provides buggy and fixed Java revisions, fault-triggering tests, and isolated bug-fix diffs \citep{just2014}. QuixBugs comprises short single-function algorithmic tasks in Java and Python \citep{lin2017}. These benchmarks support controlled experiments, whereas repository-level issues involve cross-file dependencies and specification ambiguities. A study at one granularity can illustrate a validation mechanism without showing that the same quantitative effect transfers to another.

\subsection{Additional tests provide complementary evidence}\label{sec:additional-tests-provide-complementary-evidence}

DiffTGen uses generated inputs to expose semantic differences between a candidate patch and a reference version, then relies on an oracle to judge whether a difference reveals a fault \citep{xin2017}. Its evaluation found exposing test cases for 39 of 79 possibly incorrect candidates. That denominator matters: the 79 patches were candidates requiring assessment rather than a set already adjudicated as incorrect. An unsuccessful search for a counterexample leaves the correctness judgment unresolved.

PATCH-SIM and TEST-SIM assume that passing executions should change less after a repair than failing executions \citep{xiong2018}. On a generated-patch sample, the combined heuristic filtered out 62 of 110 patches labeled incorrect while retaining all 29 labeled correct. In the same study, the heuristic rejected 16 of 194 developer patches, or 8.2\%, and documented false rejections when valid repairs altered control flow or execution traces. Full retention of the 29 correct generated patches is therefore specific to that sample, and the developer-patch results show how trace-similarity assumptions can penalize legitimate restructuring.

In a 638-patch Defects4J dataset, random testing with ground truth (RGT) identified 274 of 381 patches previously labeled overfitting \citep{ye2021rgt}. Among 257 patches originally labeled correct, generated tests raised 16 discrepancies. Follow-up analysis identified ten mistaken original labels and six false alarms. Input preconditions and imperfections or optimizations in reference patches contributed to the latter cases. These results support automatic discovery of disagreements followed by inspection of the requirements and input assumptions.

The QuixBugs journal study extended an earlier workshop report to 338 plausible patches and compared random testing, input sampling, and invariant-based assessment \citep{ye2021quix}. Its findings apply to the augmented benchmark and journal sample. Because the workshop report and journal article share this study lineage, we treat them as a single primary contribution rather than independent replications.

\begin{table}[!htbp]
\centering\small
\caption{Selected classical assessment results, with their original denominators.}
\label{tab:classical}
\begin{tabular}{@{}>{\raggedright\arraybackslash}p{.24\linewidth}>{\raggedright\arraybackslash}p{.34\linewidth}>{\raggedright\arraybackslash}p{.36\linewidth}@{}}
\toprule
\textbf{Study and setting} & \textbf{Reported observation} & \textbf{Appropriate interpretation} \\
\midrule
DiffTGen; small Defects4J patches & Exposing tests for 39/79 possibly incorrect patches. & Counterexamples were found for part of a candidate set; the remainder were unresolved by this procedure. \\
Behavioral similarity; generated-patch set & Rejected 62/110 labeled incorrect; retained 29/29 labeled correct. & Performance depends on the selected patch population and behavioral assumptions. \\
RGT; 638 Defects4J patches & Detected 274/381 previously labeled overfitting; 16 disagreements among 257 labeled correct. & Follow-up identified ten label errors and six false alarms. \\
Independent annotation study; 189 patches & Five judgments per patch from 35 developers; all-five agreement for 118 patches. & Agreement provides evidence about annotation consistency, with disagreement retained. \\
\bottomrule
\end{tabular}
\end{table}

Sources: \citep{xin2017,xiong2018,ye2021rgt,le2019}. Counts refer to different datasets and procedures and are not pooled.

\subsection{Test augmentation can change both the oracle and the search}\label{sec:test-augmentation-can-change-both-the-oracle-and-the-search}

The value of additional tests depends on which behaviors their assertions enforce. UnsatGuided adds generated tests to Nopol's repair constraints and uses reference-derived assertions separately for retrospective evaluation \citep{yu2019unsat}. In its reported analysis, regression introduction decreased in 16 of 19 relevant cases, whereas incomplete fixing improved in two cases and worsened in two. Assertions copied from the buggy program can lock in the faulty behavior that needs to change. This contrast separates preserving existing behavior from completing the intended repair.

Opad checks crashes and memory safety without requiring a reference functional output \citep{yang2017opad}. It rejected 321 of 427 patches labeled overfitting in its evaluated set, and integration into SPR produced one additional correctly repaired bug. The rejection count of 321 out of 427 measures patch filtering, whereas the single additional repaired bug measures end-to-end repair utility. The paper also documents a valid patch for the target bug that triggers an unrelated latent defect. A memory-safety failure provides concrete execution evidence, yet attributing responsibility to the candidate patch requires an explicit repair contract.

A later contract-based assessment uses Pex to search for specification violations and replays discovered counterexamples on the original-language patched methods \citep{zemin2025criteria}. Larger test suites reduce some overfitting, yet can also reduce correct-fix yield under the experimental resource limit. More test execution consumes time that the repair system could otherwise use to explore candidates. The paper's small programs and available contracts support a stronger oracle than many repository tasks offer, while bounded search and possible specification errors remain relevant.

\subsection{Labels and patch distributions require continuing scrutiny}\label{sec:labels-and-patch-distributions-require-continuing-scrutiny}

Le et al. asked 35 professional developers to assess 189 patches and collected five independent judgments per patch \citep{le2019}. All five reviewers agreed on 118 of the 189 patches, or 62.4\%, and majority ratings aligned with original author labels on most inspected subsets, while 71 patches had at least one dissenting judgment, as summarized in Table~\ref{tab:classical}. Because the reviewers were not the original project maintainers, their consensus provides an external annotation reference rather than direct access to the unwritten specification. Generated differential tests supplied complementary counterexamples and exposed label errors in several previously accepted patches.

Validator performance also shifts when the candidate patch distribution changes. Yang et al. compiled 1,988 plausible PraPR patches, including 83 labeled correct and 1,905 labeled overfitting, and reevaluated nine patch-correctness checkers \citep{yang2023}. Similarity heuristics and learned classifiers degraded on the new distribution, and dynamic tools also incurred false rejections on complex edits. A validator calibrated on small template-based edits therefore requires fresh empirical validation before certifying multi-file coding-agent patches. PRISM extends this evaluation from patch classification to ranked candidate lists and bug-level repair yield \citep{song2025prism}. Its experiments show that several filters reduce manual review volume while discarding correct top-ranked fixes.

\section{Evidence from repository-level coding agents}\label{sec:agents}

\subsection{Generated tests participate in the repair process}\label{sec:generated-tests-participate-in-the-repair-process}

Repository-level repair often begins without an executable test supplied specifically for the new issue. Existing repository tests provide regression evidence, while the future developer test is held back by benchmark construction. Agentless generates issue-reproduction tests to select candidate fixes in this setting \citep{xia2025agentless}. SWE-agent allows the system to write and run reproduction code within its trajectory \citep{yang2024sweagent}. These tests can clarify the task and expose failures; they can also share the interpretation errors that shaped the candidate patch.

When a model uses a test, failure message, coverage trace, or repeated pass/fail response to revise a patch, that information influences candidate selection. The test remains useful in the final regression record, where its use during development should be recorded explicitly. It cannot then be treated as unseen validation.

The controlled overfitting study on TDD-Bench Verified examines this issue directly \citep{ahmed2026overfitting}. It uses 449 tasks, Agentless-generated initial code, and tests produced by e-Otter++. Among initial patches that pass the generated tests, 50 of 229 for Claude-3.7-Sonnet and 58 of 176 for GPT-4o fail withheld evaluation. The corresponding conditional proportions are 21.8\% and 33.0\%. They describe disagreement between generated-test acceptance and the benchmark judge within two model configurations.

After iterative refinement, those proportions shift to 64/251 and 71/198, or 25.5\% and 35.9\%. Both the numerator and denominator change because additional patches pass the observed generated tests. This finding supports concern about adaptive test reuse under that refinement loop; it does not show that every form of test-guided refinement degrades repair quality. The same study's limit experiment exposes golden reproduction tests during repair and finds patches that pass them while failing withheld regression suites. Reproduction and regression preservation remain distinct requirements even when the reproduction test comes from the benchmark oracle.

\subsection{Broader regression tests and differential tests expose different failures}\label{sec:broader-regression-tests-and-differential-tests-expose-different-failures}

PatchDiff audits 877 plausible patches produced by three systems on SWE-bench Verified \citep{wang2026patchdiff}. Executing all available developer tests identifies 68 functionally incorrect patches. Generated differential tests flag 260 of the 877 patches as having suspicious candidate-reference discrepancies; 215 had not been identified by the broader developer-test check. The procedures therefore detect partly different behaviors.

UTBoost audits the evaluation machinery and separates test-log parser repairs from newly generated test cases \citep{yu2025utboost}. Parser changes alone alter 64 Lite and 79 Verified patch outcomes in the reported experiments; the benchmark subsets overlap. A recorded verdict can thus misrepresent the test run itself. Added tests address a separate limitation: the set of behaviors executed.

PatchDiff's manual assessment of 77 sampled suspicious patches assigns 22 to incorrect, four to correct, and 51 to uncertain. A differential test may uncover an input whose intended outcome is unclear or a behavior introduced incidentally by the reference patch. Treating every discrepancy as an error would bypass the adjudication required by the study.

PatchDiff's headline estimate of resolution-rate inflation combines the incorrectness rate in the 77-patch manual sample with errors detected outside the discrepancy set \citep{wang2026patchdiff}. We use the directly reported counts of 68/877 broader-test failures, 260/877 differential discrepancies, and 22 incorrect, four correct, and 51 uncertain cases in the 77-patch sample to show complementary detection and adjudication uncertainty. Extrapolating those sample proportions to all 877 patches remains conditional on the study's sampling assumptions.

\subsection{Reproducing a reference transition and discriminating patches}\label{sec:reproducing-a-reference-transition-and-discriminating-patches}

SWT-Bench evaluates issue-reproduction tests and studies their use in filtering generated fixes \citep{mundler2024swtbench}. In the reported filtering experiment, precision reaches 47.8\% while recall is 20\%. Both values are needed: a selective filter can increase the proportion of accepted patches that satisfy the benchmark while discarding many successful candidates. The paper explicitly limits its benchmark to bug reproduction and identifies edge-case detection and global coverage as additional concerns.

SWEPatchTest evaluates generated tests against pools of candidate patches as well as the buggy and reference versions \citep{lu2026swepatchtest}. The inspected preprint contains 619 tasks and 82,579 non-reference patches. It asks whether a test suite reproduces the reference transition, accepts the sampled patches labeled correct, and rejects the sampled patches labeled incorrect. Its joint validation criterion requires these conditions on the same task.

The highest reported main-configuration joint validation rate is 9.21\%; the same configuration reproduces the reference transition on 59.29\% of tasks. Joint validation also requires agreement for every sampled candidate, so its success rate depends on the candidate pool's size, composition, and sampling rules. Successful reference reproduction alone often fails to meet this stricter criterion.

The candidate labels introduce a further dependency. SWEPatchTest assigns them by executing each source benchmark's official oracle. Its evaluation measures agreement with those operational labels. An incorrect patch accepted by the source oracle can enter the positive pool, and a valid alternative rejected by that oracle can enter the negative pool. The paper acknowledges this boundary. The benchmark expands the set of implementations tested against a generated suite, while the semantic meaning of its labels continues to depend on the underlying judge.

\begin{table}[!htbp]
\centering\small
\caption{Repository-level evidence and the limits of each numerical comparison.}
\label{tab:repo}
\begin{tabular}{@{}>{\raggedright\arraybackslash}p{.24\linewidth}>{\raggedright\arraybackslash}p{.34\linewidth}>{\raggedright\arraybackslash}p{.36\linewidth}@{}}
\toprule
\textbf{Study} & \textbf{Sample and observation} & \textbf{Boundary of the inference} \\
\midrule
Generated-test overfitting & 449 tasks; hidden-evaluation failures among generated-test-passing initial patches: 50/229 and 58/176. & Conditional disagreement for two models and one repair/test-generation setup. \\
PatchDiff audit & 68/877 fail broader developer tests; 260/877 show differential discrepancies; sampled discrepancies: 22 incorrect, 4 correct, 51 uncertain. & Discrepancy and adjudicated incorrectness have separate labels; manual sample is drawn from suspicious patches. \\
SWT-Bench patch filtering & Reported precision 47.8\%, recall 20\%; raw filtering counts are not reported in the inspected passage. & Precision reaches 47.8\%, while 20\% recall discards four-fifths of benchmark-passing patches. \\
SWEPatchTest & 619 tasks; same configuration gives 59.29\% reference reproduction and 9.21\% joint validation. & Joint success depends on candidate pools and official-oracle labels; preprint evidence. \\
\bottomrule
\end{tabular}
\end{table}

Sources: \citep{ahmed2026overfitting,wang2026patchdiff,mundler2024swtbench,lu2026swepatchtest}. The studies use distinct denominators and outcomes.

\subsection{Specification quality affects both acceptance and rejection}\label{sec:specification-quality-affects-both-acceptance-and-rejection}

Task descriptions and hidden tests can disagree about what a solver was expected to know. Agentless reports an inspection of the 300 SWE-bench Lite tasks in which 10.0\% lacked sufficient issue-description information, including required names or exact strings absent from the stated issue \citep{xia2025agentless}. This is evidence about that benchmark subset and the authors' assessment procedure. It motivates examining whether a failure reflects a defective patch or an under-specified task.

SWE-Bench Pro augments task descriptions with requirements and interfaces to improve resolvability, and its limitations acknowledge that test suites can reject valid alternative implementations \citep{deng2026swebenchpro}. More explicit requirements can reduce ambiguity. They also change the task presented to the agent, so their origin and timing should be reported. A benchmark may evaluate resolution from an original issue or resolution from a curated specification; these are distinguishable operating conditions.

Function-level benchmarks provide a supporting comparison under simpler state assumptions. EvalPlus expands test suites with model-generated and mutated inputs, using precondition contracts to filter valid inputs and reference implementations to supply expected outputs \citep{liu2023evalplus}. Those expanded suites expose failures missed by original HumanEval tests. This result illustrates how stronger input sampling tightens validation, though its quantitative effect remains tied to self-contained functions. Repository-level maintenance requires additional checks for shared state, external dependencies, and regression scope.

\subsection{Training exposure and test adequacy require separate checks}\label{sec:training-exposure-and-test-adequacy-require-separate-checks}

Data contamination concerns whether evaluation information was available during training or other prior adaptation. Test-suite overfitting concerns whether acceptance under observed checks generalizes to additional requirements. A fresh task can have weak tests, and a strong test suite can evaluate a task whose solution appeared in training. Combining these problems into one validity label obscures the appropriate remedy \citep{renzullo2025}.

LiveCodeBench uses timestamped competition tasks to support evaluation windows relative to model cutoffs \citep{jain2025livecodebench}. SWE-bench Goes Live extends time-aware collection to repository issues and provides task-specific executable environments \citep{zhang2025swebenchlive}. These approaches improve control over task provenance and execution. Comparing scores across an older and a newer benchmark also changes the sampled tasks, so a score difference alone does not isolate the effect of memorization.

BenchChecker probes repository presence and patch persistence to investigate contamination in bug-fixing benchmarks \citep{wu2026benchchecker}. Its assumptions include the interpretation of model outputs, the availability of cutoff information, and whether relevant patches persisted in repository snapshots. Such diagnostics provide evidence about exposure. They should be reported with those assumptions, separately from the tests used to judge candidate behavior.

\subsection{Strengthened tests can also reject valid alternatives}\label{sec:strengthened-tests-can-also-reject-valid-alternatives}

SWE-ABS uses reference-aware generation, coverage, and adversarial mutations to strengthen benchmark tests \citep{yu2026sweabs}. Before reporting the main evaluation, its authors inspected all 500 Verified tasks and identified 53 with new tests tied to particular reference-patch behavior. They revised or removed those tests. The fraction 53/500 describes tasks containing problematic added tests; it is distinct from the proportion of valid candidate patches rejected by a validator.

After revising the tests, the paper reports 2,184 rejected submissions among 11,041 previously passing patches in Table 2. A separate manual study classifies 100 rejected patches. The inspection of every task concerns test design, whereas the manual sample concerns rejected candidates. The larger rejection count depends on the revised suite and its oracle assumptions. The abstract and main table give different rounded rate summaries, so the counts are retained here.

An institutional audit by OpenAI provides further evidence about specification-test mismatch \citep{openai2026audit}. It selected 138 of the 500 Verified tasks using inconsistent model performance across repeated runs and assigned at least six experienced-engineer reviews per selected task. The report identifies implementation-specific tests and requirements absent from the issue description, alongside exposure probes. Its failure-conditioned sample is informative about mechanisms that frustrate evaluation; prevalence estimates for the complete benchmark would require a representative design. This non-peer-reviewed report is treated as a complementary source, with its selection procedure and institutional provenance retained.

A permissive oracle can accept faulty code, while a restrictive one can reward resemblance to an undisclosed reference implementation. Oracle strengthening should therefore be assessed for both its detection of invalid candidates and its retention of admissible alternatives, especially when test generation uses the reference patch.

\section{Comparative case audits and worked applications}\label{sec:synthesis}

\subsection{Compare evaluation decisions under explicit conditions}\label{sec:compare-evaluation-decisions-under-explicit-conditions}

A reported score combines a candidate population with a decision rule for accepting, rejecting, or leaving a patch unresolved. The four comparative audits in Table~\ref{tab:comparisons} apply our five-dimensional profile to evaluation decisions that look similar when reduced to a single headline rate. The archived evidence matrix records the same five dimensions across all 33 primary contributions.

When candidate patches and deterministic test executions are held fixed, adding tests that must all pass can only preserve or reduce the set of accepted patches: any patch passing the enlarged suite also passes its initial subset. Whether newly rejected patches are genuinely faulty depends on the added tests' oracle assumptions. Rerunning repair search with the enlarged suite introduces a second effect: additional tests can redirect candidate generation, rule out valid fixes, or consume the execution budget. Separating post-hoc filtering from in-search test augmentation explains the contrasting outcomes across test-augmentation studies \citep{yu2019unsat,zemin2025criteria}.

\begin{table}[!htbp]
\centering\small
\caption{Conditions governing four comparative audits across the 33-study dataset.}
\label{tab:comparisons}
\begin{tabular}{@{}>{\raggedright\arraybackslash}p{.24\linewidth}>{\raggedright\arraybackslash}p{.34\linewidth}>{\raggedright\arraybackslash}p{.36\linewidth}@{}}
\toprule
\textbf{Comparison} & \textbf{Conditions that change} & \textbf{Defensible interpretation} \\
\midrule
Le et al. versus RGT test detection & Patch populations, label sources, denominator, generation settings, and observation rules. & Detection differs across configured procedures; the headline counts do not isolate a causal effect of test independence. \\
SWT-Bench versus SWEPatchTest & Test-quality target, task collection, candidate pools, and correctness-label construction. & Reference reproduction and all-candidate discrimination test different claims. \\
PatchDiff detection versus human judgment & Full patch set versus discrepancy-enriched sample; test failure versus requirement judgment. & Automatic discrepancy identifies cases for adjudication, with 51 of 77 inspected cases remaining uncertain. \\
Original PATCH-SIM, later reassessment, and PRISM utility study & Candidate distribution, supported subset, operating rule, and patch-versus-bug endpoint. & Error detection and preservation of useful fixes must be assessed together under the intended selection policy. \\
\bottomrule
\end{tabular}
\end{table}

\subsection{Why independent-testing results differ}\label{sec:why-independent-testing-results-differ}

Le et al. report 27 detections by DiffTGen and 31 by Randoop, with a union of 51, across 189 sampled patches \citep{le2019}. RGT reports 274 detections among 381 patches previously labeled overfitting \citep{ye2021rgt}. Dividing 51 by 189 and comparing it directly with 274/381 would mix a full sample containing several label categories with an incorrect-only denominator. The independent annotation in Le and the inherited labels plus targeted reassessment in RGT also supply different reference judgments.

RGT also compares procedures on the same 381 labeled-overfitting patches: its combined generators identify 274 and its configured DiffTGen identifies 143. The patch set is fixed, while several implementation choices differ. The reported EvoSuite search budgets are 100 and 60 seconds; target construction, instrumentation, and handling of execution differences also change. The counts describe the configured procedures. Attributing the difference to one component would require an ablation holding the other conditions fixed.

The paper's case analyses explain some detection differences. In its reassessment of patches originally labeled correct, two Lang-43 examples encounter a heap-space failure that leaves DiffTGen without instrumented values to compare, while RGT can flag the execution discrepancy. Those examples come from a different label stratum than the 381-patch comparison. Other examples involve failure to generate a distinguishing input. The cases identify mechanisms involving observation and input reachability, although their respective contributions to the overall detection difference remain unmeasured.

The two studies operate on different candidate sets, reference labels, execution budgets, and rules for ambiguous outcomes. Their headline detection rates, 51/189 across mixed labels in Le et al. versus 274/381 among overfitting-labeled patches in RGT, therefore answer different questions. Isolating the effect of a single test generator requires holding the candidate pool, budget, and oracle rules fixed.

\subsection{Why reproduction and discrimination scores diverge}\label{sec:why-reproduction-and-discrimination-scores-diverge}

SWT-Bench tests whether a generated suite distinguishes the buggy and reference versions, with modified-code coverage as additional information \citep{mundler2024swtbench}. SWEPatchTest adds candidate pools whose members are labeled through the source benchmarks' official tests \citep{lu2026swepatchtest}. A jointly validated task must reproduce the reference transition, accept every sampled positively labeled candidate, and reject every sampled negatively labeled candidate.

By construction, jointly validated tasks form a subset of the tasks that satisfy reference reproduction. On a fixed task set, expanding a candidate pool can only preserve or reduce the all-candidate pass rate when the test suite, candidate labels, execution outcomes, and error-handling rules remain fixed. Resampling candidates or re-executing flaky tests alters those conditions, which is why a joint validation rate requires the pool size and composition alongside it. The reported 59.29\% reference reproduction and 9.21\% joint validation for one configuration in SWEPatchTest quantify the added candidate-discrimination constraints within that study \citep{lu2026swepatchtest}; they do not measure a temporal decline from SWT-Bench \citep{mundler2024swtbench}.

Two oracle layers affect this comparison: the generated suite judges candidates whose labels were assigned by another test oracle. Agreement with those labels measures discrimination relative to that operational standard. The benchmark retains unresolved executions in the denominator and awards them neither acceptance nor rejection credit, so joint success also requires interpretable execution outcomes. An independently audited subset would be needed to determine how often label errors alter the conclusion. Reports should distinguish agreement with benchmark labels from agreement with requirements.

\subsection{Reconstructing a repository-level audit}\label{sec:reconstructing-a-repository-level-audit}

\begin{figure}[!htbp]
\centering
\includegraphics[width=\linewidth]{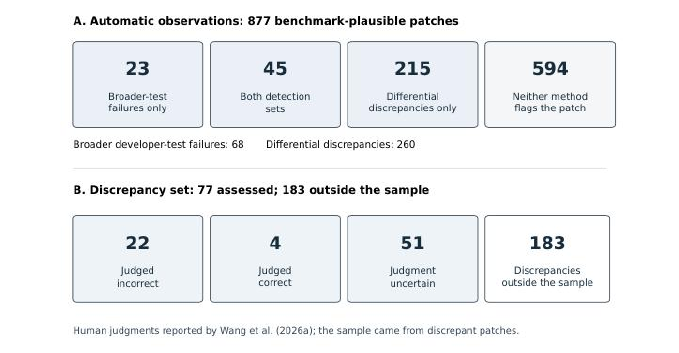}
\caption{Counts from the PatchDiff audit \citep{wang2026patchdiff}, Sections 4.1, 4.2, and 4.4. The two detection sets overlap: 45 patches are in both, 23 only in broader-test failures, 215 only in differential discrepancies, and 594 in neither set. These four values partition the 877 candidates. The source authors' human assessments cover a 77-patch sample from the 260 discrepancies. The accompanying figure data identify counts derived by subtraction.}
\label{fig:patchdiff}
\end{figure}

Figure~\ref{fig:patchdiff} applies the profile to PatchDiff's reported observations \citep{wang2026patchdiff}. Among 877 plausible patches, broader developer tests flag 68 and differential testing flags 260. The paper reports 215 differential discrepancies outside the developer-test failures. The resulting intersection is 45, derived as 260 minus 215. This overlap is needed to avoid counting the same patch twice.

PatchDiff's audit starts with archived, already accepted candidates and accesses the developer reference patch after repair \citep{wang2026patchdiff}. Its differential tests therefore provide retrospective audit evidence rather than checks available during live issue resolution. The execution rule is asymmetric: each version runs 20 times, and a discrepancy is recorded when one version passes all 20 runs while the other fails at least once. This rule requires stability on the passing side while counting a single failure on the other; reporting only the repetition count omits the acceptance criterion.

Of the 77 discrepant patches manually inspected by the PatchDiff authors, 22 are judged incorrect, four correct, and 51 uncertain \citep{wang2026patchdiff}. The remaining 183 discrepant patches fall outside that manual sample, and the 594 patches flagged by neither automatic check also require further evidence before a semantic correctness claim can be made. The evidence matrix records automatic detection outcomes separately from manual semantic judgments. This separation distinguishes an observed discrepancy from an adjudicated fault or an unresolved case.

\subsection{From patch classification to useful repairs}\label{sec:from-patch-classification-to-useful-repairs}

The original PATCH-SIM evaluation rejects none of 29 correct generated patches, while a separate developer-patch evaluation in the same paper rejects 16 of 194 \citep{xiong2018}. The later reassessment reports three rejections among ten correct patches in the method's supported subset, alongside 210 rejections among 493 overfitting patches \citep{yang2023}. Different populations and support exclusions account for different denominators. The original paper's developer-patch result already identifies the vulnerability of the trace-preservation assumption to legitimate behavioral restructuring.

PRISM evaluates how filtering affects repair selection \citep{song2025prism}. Across ten repair tools, the number of correct top-1 tool/bug outcomes decreases from 176 to 127 under the studied PATCH-SIM filter and increases to 185 under PRISM. The same bug can contribute an outcome for more than one tool. PRISM also reduces the sum of first-fix candidate inspections across tools from 609 to 537 in that setting. A filter's ability to reject wrong patches therefore needs to be assessed alongside its retention of the correct candidates available for each bug.

Incorrect-patch detection uses all adjudicated incorrect candidates as its denominator; correct-patch retention uses all adjudicated correct candidates. Bug-level yield also depends on ranking, stopping, and inspection cost. Rejecting one correct patch can have little effect when an equivalent correct candidate remains near the top, or eliminate the only solution found under the budget. Patch-level classification and bug-level repair success therefore require separate results.

\section{A protocol for reporting correctness evidence}\label{sec:protocol}

\subsection{Define the decision and preserve the evidence boundary}\label{sec:define-the-decision-and-preserve-the-evidence-boundary}

An evaluation should first name its decision target: issue reproduction, passage of specified regression tests, rejection of known faulty candidates, or adjudicated satisfaction of maintenance requirements. A study that addresses multiple targets should report a separate denominator and outcome for each. The protocol in Table~\ref{tab:protocol} synthesizes reporting requirements from the 33 primary contributions; its effect on downstream maintenance decisions requires empirical evaluation.

Before repair begins, the evaluation should identify the repository revision, issue text, allowed external information, and available tests. The record should distinguish tests created during repair from tests reserved for final assessment. For each interaction, it should specify what the system received: full test source, an assertion, a failure trace, coverage, or a binary verdict. Repeated verdicts can influence a search even when test source remains hidden. Feedback rounds and candidate counts document this exposure.

The final assessment should preserve its own execution record. At minimum, that record should identify which tests ran, the program and dependency versions, and how unsuccessful execution was handled. A timeout, infrastructure error, skipped assertion, and behavioral failure carry different meanings. Excluding unsuccessful executions changes the evaluated population. A study should state its denominator policy and provide the associated counts.

\subsection{Report complementary outcomes without collapsing them}\label{sec:report-complementary-outcomes-without-collapsing-them}

A repair evaluation can retain a sequence of outcomes for the same candidate: reproduction result, regression result, result on held-out tests, differential discrepancies, and final adjudication. Reporting this sequence shows what each validation stage contributes. The sequence also reveals when additional validation rejects a previously accepted patch or when a suspected discrepancy is resolved as a valid alternative.

When evaluating a test generator, the target is the quality of the test as a judge. A useful design evaluates both acceptance of valid alternatives and rejection of invalid alternatives. The SWEPatchTest construction makes this distinction operational, while its inherited labels show why the label source must accompany the result \citep{lu2026swepatchtest}. A candidate pool should include semantic diversity, and its construction should be stated explicitly. Multiple textual variants of the same behavior provide less information about a test's discrimination than behaviorally distinct candidates.

Accuracy can obscure the error type that matters in a patch filter. An evaluator that rejects most candidates may look successful in a dataset dominated by incorrect patches while losing useful repairs. The PraPR reassessment illustrates the relevance of class balance and patch-distribution change \citep{yang2023}. A report should provide the confusion counts where labels are available, the retention of correct candidates, the detection of incorrect candidates, and the number left uncertain. The uncertain group should remain visible instead of being silently assigned to a binary category.

\begin{table}[!htbp]
\centering\small
\caption{Proposed minimum record for interpreting a repair result.}
\label{tab:protocol}
\begin{tabular}{@{}>{\raggedright\arraybackslash}p{.24\linewidth}>{\raggedright\arraybackslash}p{.34\linewidth}>{\raggedright\arraybackslash}p{.36\linewidth}@{}}
\toprule
\textbf{Record} & \textbf{Required content} & \textbf{Result enabled} \\
\midrule
Task and exposure & Issue and repository identifiers; requirement amendments; model version; known training cutoff; visible test and feedback history. & Defines the task the system actually attempted. \\
Candidate production & Number of candidates; selection and stopping rule; token, time, and execution budgets. & Distinguishes one-attempt success from search-assisted selection. \\
Test inventory & Reproduction and regression suites; generated and held-out tests; input constraints; oracle source. & Links each pass or failure to an explicit behavioral claim. \\
Execution record & Environment and harness versions; commands; per-test status; retries; skipped and unresolved executions. & Supports reconstruction of the acceptance decision. \\
Adjudication & Counterexample; requirement reference; decision and confidence; disagreements; uncertain labels. & Makes semantic judgments traceable. \\
Aggregate results & Denominators at every filter; counts by outcome; repeated-run variation; validation cost. & Supports comparisons within a shared evaluation setting. \\
\bottomrule
\end{tabular}
\end{table}

\subsection{Separate benchmark auditing from prospective repair}\label{sec:separate-benchmark-auditing-from-prospective-repair}

A retrospective benchmark supplies a historical developer fix. That reference patch supports differential testing in DiffTGen, RGT, and PatchDiff \citep{xin2017,ye2021rgt,wang2026patchdiff}. During live software maintenance on an unresolved issue, no reference fix exists. A deployable validator must therefore specify what replaces reference behavior: explicit contracts, requirements-derived tests, preserved regression invariants, or maintainer review. Accuracy measured with access to a ground-truth fix should be reported explicitly as retrospective benchmark auditing.

Evaluations can compare two modes on the same task set. Benchmark auditing uses the reference patch to diagnose weaknesses in a test oracle. Prospective validation uses information available when the issue was opened. Their comparison would estimate how much assessment quality depends on hindsight and help identify information leakage through reference-derived tests.

\subsection{Make validation cost part of the result}\label{sec:make-validation-cost-part-of-the-result}

Broader testing and human adjudication require time. A practical assessment should report where that time is spent and what additional faults each stage discovers. Running a full regression suite may be expensive for a large repository; generated differential tests can require model calls and repeated execution; ambiguous discrepancies may need maintainer attention. These costs determine whether a validation policy can be used on every patch or only on selected high-risk changes.

An incremental analysis can measure newly detected faults after each stage while tracking correct candidates lost and unresolved cases added. The resulting record supports a policy decision under a specified budget. A single aggregate accuracy or resolution rate omits this trade-off. Thresholds and stopping policies should be chosen for an explicit use case, then evaluated on tasks held out from that choice.

\section{Research opportunities}\label{sec:opportunities}

\subsection{Measure the contribution of independent evidence}\label{sec:measure-the-contribution-of-independent-evidence}

The 33 primary contributions show that additional tests can expose missed faults, with outcomes that depend on the test generator, oracle source, and candidate patch distribution. A controlled follow-up experiment should hold the repair agent and candidate pool fixed while adding validation stages sequentially: reproduction tests, repository regression suites, independently generated tests, and maintainer adjudication. That design measures marginal fault detection and false rejection at each stage without confounding validator changes with a different patch generator.

Such an experiment should stratify candidates by patch size, modified component, repository, and fault category. Those strata separate failures to reach changed code from failures to judge the resulting behavior. Recording generation, execution, and review time at each stage identifies which check first exposes each adjudicated fault under a practical maintenance budget.

\subsection{Study shared interpretation errors in code and tests}\label{sec:study-shared-interpretation-errors-in-code-and-tests}

The generated-test overfitting results motivate experiments on dependence between the patch generator and the test generator \citep{ahmed2026overfitting}. A controlled study could vary the generating model, prompt, access to candidate code, and source of requirement information while keeping the task set fixed. Diversity of model names alone would be a weak operational definition of independence because models may share training sources or common interpretations. Observable disagreements and requirements-linked counterexamples provide a more concrete outcome.

The study should include test-first and patch-first conditions and record any test revisions made after observing candidate outputs. An assertion revised to accommodate a candidate supplies different evidence from one fixed before patch generation. An audited subset would allow agreement with existing benchmark labels to be compared with agreement with requirements.

\subsection{Build datasets of admissible alternatives and disputed cases}\label{sec:build-datasets-of-admissible-alternatives-and-disputed-cases}

Patch-discrimination datasets need both invalid candidates and valid alternatives. Reference-only comparisons undersample the space of correct behavior. Future datasets could curate several accepted implementations per issue, together with explicit input domains and evidence for permissibility. Maintainer feedback would be especially useful where the issue leaves choices unspecified.

Disputed cases should be retained alongside the reasons for disagreement. A dispute can reflect an underspecified issue description, an invalid generated input, a flaky test environment, or an incidental choice in the reference patch. The 51/77 uncertain cases in PatchDiff and the 6/16 false alarms in RGT show why unresolved and disputed records belong in benchmark releases \citep{wang2026patchdiff,ye2021rgt}. Publishing only clear-cut examples underrepresents the maintenance decisions that require human review.

\subsection{Evaluate transfer across generators and time}\label{sec:evaluate-transfer-across-generators-and-time}

The classical reevaluation of correctness checkers shows sensitivity to candidate distribution \citep{yang2023}. Repository agents can change that distribution through new models, longer searches, and multi-file edits. A future benchmark should therefore split assessment data by repository, time, and patch generator, and report each split's assumptions. Reusing random patch-level splits can place closely related edits on both sides of an evaluation.

Refreshed tasks and controlled private tasks offer complementary ways to study exposure, while reproducible environments support execution comparisons \citep{zhang2025swebenchlive,deng2026swebenchpro}. Longitudinal studies should retain previous benchmark releases and rerun a stable subset of systems to separate changes in the task population from changes in the repair system. Where a closed model changes without a stable version identifier, the resulting version uncertainty should be reported as a limitation of the longitudinal comparison.

\section{Threats to validity}\label{sec:threats}

Targeted query construction and artifact selection introduce sampling boundaries. Search ranking, open-access availability, language, and the focus on widely used benchmarks shape the 39-source dataset. Our findings concern validation mechanisms and empirical observations documented in the audited studies rather than a population prevalence across all APR research. The archived evidence matrix and query log document every selection decision for verification and extension.

Passage extraction and coding across heterogeneous studies can omit qualifications, merge versions, or conflate task counts with patch counts. Section and table locators, denominator definitions, version notes, and transfer boundaries accompany each coded record to support independent verification. The supplementary candidate ledger documents the 57-record coverage check, and the authors verified all cited literature, numerical denominators, and cross-study comparisons against the primary texts.

Construct validity depends on how each primary study defines correctness. Held-out test suites, reference implementations, external developer panels, and project maintainer reviews rest on different assumptions. Because several studies reuse Defects4J, QuixBugs, or SWE-bench subsets, their findings are not statistically independent replications. This study preserves each benchmark's operational definition and avoids pooling incompatible rates. Our five-dimensional profile is an analytical framework whose inter-annotator agreement remains to be measured.

External validity is bounded by the tasks studied. Small C programs, Java benchmark defects, Python repository issues, and programming-contest functions differ in scale and context. Evidence from one setting is used to explain a mechanism; quantitative effects remain attached to that setting. Closed-model behavior, evaluator configurations, and benchmark revisions introduce further limits on transfer. Recent preprints contribute relevant observations, with their publication status explicitly retained.

Our audit analyzes reported execution records and benchmark artifacts. Repair systems and test generators were not rerun, so execution reproducibility remains bounded by the primary studies unless directly checked in their released artifacts. The reporting protocol, staged validation experiments, and dataset designs proposed here require prospective empirical evaluation.

\section{Conclusions}\label{sec:conclusions}

Evaluating patch correctness across classical repair systems and coding agents requires tracing where tests originated, which behaviors and oracles they check, how they executed, and how discrepancies were adjudicated. Classical APR studies showed that test-suite passage can conceal both behavioral overfitting and test-harness defects \citep{smith2015,qi2015}. Repository-level coding agents generate reproduction tests and consume iterative execution feedback during repair, blurring the boundary between search-visible checks and independent final assessment. Broader regression suites and differential testing expose complementary faults, while behavioral discrepancies against a reference fix still require adjudication against the issue's stated requirements.

Our four worked comparative audits show why a single resolution or detection rate cannot by itself guide benchmark or maintenance decisions. Changed candidate denominators, reference-dependent labels, overlapping detection sets, and the gap between patch-level filtering and bug-level repair yield all alter the meaning of reported scores. Recording incorrect-patch detection, correct-patch retention, unresolved cases, and validation cost together allows researchers and maintainers to judge when additional testing or human review justifies its cost.

\FloatBarrier
\section*{Declarations}

\noindent\textbf{Funding.} This research received no external funding.

\medskip
\noindent\textbf{Competing interests.} The authors declare no competing financial interests or personal relationships that could have influenced the work reported in this article.

\medskip
\noindent\textbf{Author contributions.} Shuai Wu: Conceptualization, Methodology, Investigation, Data curation, Writing - original draft, Writing - review \& editing. Yanna Feng: Conceptualization, Methodology, Investigation, Validation, Visualization, Supervision, Writing - original draft, Writing - review \& editing. Yufang Li, Xue Li, Zhijun Wang, Bolun Liu, and Weilin Cai: Investigation, Validation, Data curation, Writing - review \& editing.

\medskip
\noindent\textbf{Data availability.} The source inventory, five-dimensional evidence matrix, extracted claims, comparison cases, supplementary candidate ledger, search log, figure data, and bibliographic records are publicly archived in Zenodo as version 1.0. The dataset is described by \citet{wu2026reviewdata} and is available at \url{https://doi.org/10.5281/zenodo.23639376}.

\appendix
\section{Coding rules and evidence records}\label{app:coding}

The archived evidence matrix contains one row for each of the 33 primary contributions, comprising 32 primary studies and one institutional technical audit. Each row records the study design, candidate population, unit of analysis, five evidence-profile dimensions, supported claim, transfer boundary, and benchmark or label ancestry. Every profile dimension includes a descriptive category, a passage-level summary, and a section or table locator. The six taxonomic baselines appear in the bibliographic inventory and Table~\ref{tab:reviews} without being counted as primary observations.

For test provenance and exposure, we record which component could access a test, reference patch, assertion, or execution outcome at each stage of repair and evaluation. A test whose source code is hidden but whose pass/fail verdict guides candidate selection is classified as search-visible evidence. Procedural separation holds only when both test content and execution outcomes remain withheld until candidate selection is complete. Random-seed separation in test generation is recorded separately from oracle independence.

For behavioral scope, we record the checked property: issue reproduction, regression preservation, reference agreement on generated inputs, crash and memory safety, or discrimination across candidate pools. Code coverage records whether execution reached modified statements, whereas the expected-outcome rule is recorded under oracle basis. Syntactic similarity, execution-trace similarity, and learned representations are recorded as behavioral proxies together with their structural assumptions.

For oracle basis and dependence, we record two sources when applicable: the runtime oracle that issues a verdict during validation, and the evaluation reference that labels each verdict as right or wrong. A validator can run without a developer patch while still being benchmarked against reference-derived labels. Shared authors, institutions, benchmarks, and inherited label sets are recorded as provenance and ancestry links. Neither relationship supplies an exact count of overlapping patches or quantifies the statistical dependence of individual observations.

For execution integrity, we record the execution environment, test-selection and log-parsing rules, resource budgets, handling of timeouts and errors, and repeated runs. Whenever a study uses repeated executions, we record its multi-run acceptance rule alongside the repetition count. Evidence reported for one experimental subset is not extrapolated to other subsets in the same paper, and unresolved reporting ambiguities are noted explicitly in the matrix.

For label adjudication, we record the evaluator, inspected artifacts, decision categories, and sampled population. Full-benchmark task audits and sampled candidate-patch audits are recorded as distinct procedures. An automated execution discrepancy remains recorded as a discrepancy unless the primary study reports a semantic correctness judgment, and uncertain verdicts retain their own category. Counterexample inputs are interpreted relative to the preconditions stated in the source study.

``Not established in inspected passages'' marks a field for which the retrieved material does not support a stronger statement. It is distinct from a documented absence and from a dimension that is outside a study's endpoint. The fields are descriptive and are not summed into a quality score. The authors verified the extracted passages, coding entries, cited literature, key numerical results, and cross-study comparisons. The coding instrument's inter-annotator reliability and prospective empirical validity remain to be measured.

\FloatBarrier
\bibliographystyle{unsrtnat}
\bibliography{references}
\end{document}